# Pressure-Induced Superconductivity and Its Scaling with Doping-Induced Superconductivity in the Iron Pnictide with Skutterudite Intermediary Layers

Peiwen Gao[1], Liling Sun[1]*, Ni Ni[2,3,4], Jing Guo[1], Qi Wu[1], Chao Zhang[1], Dachun Gu[1], Ke Yang[5], Aiguo Li[5], Sheng Jiang[5], Robert Joseph Cava[2] and Zhongxian Zhao[1]*

[1]Institute of Physics and Beijing National Laboratory for Condensed Matter Physics, Chinese Academy of Sciences, Beijing 100190, China

[2]Department of Chemistry, Princeton University, Princeton, NJ 08544, USA

[3]Los Alamos National Laboratory, Los Alamos, NM87544, USA

[4]Department of Physics and Astronomy, UCLA, Los Angeles, CA90095, USA

[5]Shanghai Synchrotron Radiation Facilities, Shanghai Institute of Applied Physics, Chinese Academy of Sciences, Shanghai 201204, China

E-mail: llsun@iphy.ac.cn

The discovery of superconductivity near 26 K in the iron pnictide $LaFeAsO_{0.9}F_{0.1}$[1] manifested the birth of the second family of high temperature superconductors after the first family of copper oxide superconductors was discovered. Since then, a great deal of fruitful efforts has been made in searching for the other new iron-based superconductors. The $Ca_{10}(Pt_nAs_8)(Fe_2As_2)_5$ (n=3, 4) compounds are new type of iron pnictide superconductors[2-8] whose structures can be described as replacing alternatively the $Fe_2As_2$ layers in the $CaFe_2As_2$ (Ca-122) unit cell with $Pt_nAs_8$ intermediary layers (so called skutterudite layer), *i.e.*, by stacking Ca-$Pt_nAs_8$-Ca-$Fe_2As_2$ layers in a unit cell. The two different types of intermediary layers have a diverse influence on the physical properties. When they are $Pt_3As_8$, which have a semiconducting nature, the undoped $Ca_{10}(Pt_3As_8)(Fe_2As_2)_5$ (the "10-3-8 phase") compound is an antiferromagnetic semiconductor and non-superconducting down to 1.5 K at ambient pressure. Upon chemical doping with Pt on the Fe site of the FeAs layer in the 10-3-8 compound, the system can be tuned into the superconducting state.[2-5,8] While, when the intermediary layers are $Pt_4As_8$, which have a metallic nature, the 10-4-8 compound presents metallic behavior at ambient pressure as most of the other iron pnictide superconductors and shows superconductivity at 26 K. [2,4,5] Apparently, the 10-3-8 phase is special due to its unique intermediary layers with semiconducting character, no other examples of this are known in the pnictide superconducting literature. As the 10-3-8 phase is made of both charge reservoir layers and electronically active layers, it nicely connects the iron pnictide superconductors with the copper oxide superconductors.

It is known that high temperature superconductivity usually develops in the proximity of a suppressed antiferromagnetic (AFM) ordered phase that can be tuned by control parameters such as pressure or carrier doping. [9-11] In practice, high pressure can serve dual roles in tuning both the stability of AFM long-range order and the state of carrier density at the Fermi surface through adjustment of the lattice parameters. Therefore, it is commonly adopted as an important tool to explore new superconductors and new phenomena in the known superconductors. Here we report systematic high-pressure studies on the undoped semiconducting 10-3-8 compound through a combination of *in-situ* resistance, *ac* susceptibility, Hall coefficient and synchrotron X-ray diffraction measurements performed with diamond anvil cells. We find the emergence of superconductivity in the 10-3-8 phase under pressure after the long-range AFM order is completely suppressed. The temperature-pressure phase diagram obtained shows that a superconducting dome with maximum $Tc$ of 8.5 K at 4.1 GPa lies in the pressure range of 3.5-7 GPa. No pressure-induced structural phase transitions are observed in the 10-3-8 compound for the pressure range investigated at room temperature. Scaling the changes between the pressure-induced and the doping-induced superconductivity in this system reveals obvious similarities and dramatic differences, demonstrating the presence of different mechanisms of developing and stabilizing superconductivity in these two cases. Equally important, we find that the pressure/doping-temperature phase diagram of the 10-3-8 phase is different from that of other known iron pnictide (e.g. 122-type) systems, which present a coexistence region of AFM phase and superconducting phase. [12,13] These experimental results may provide important information for further investigations of both theoretical and experimental aspects of superconductivity in the iron-based superconductors.

As shown in Fig.1a, the diffraction patterns perpendicular to the crystal plates show only narrow (00*l*)-type reflections, indicating that the crystals are high quality and single phase. Chemical analysis (employing the WDS method) shows that the composition of the crystals studied is $Ca_{10}(Pt_3As_8)(Fe_{1.996}Pt_{0.004}As_2)_5$.[3] Since the amount of Pt in the $Fe_2As_2$ layers is much too small to affect either the physical or structural properties, we denote the material as undoped 10-3-8 hereafter.

The temperature ($T$) dependent resistance ($R$) of the 10-3-8 phase measured at ambient pressure displays three significant features (Fig.1b). The first is an obvious resistance minimum near 210 K; the resistance decreases with cooling and then increases with further cooling, demonstrating a crossover from a metallic to a semiconducting state (Here we define the crossover temperature as $T'$). The other two features can be seen by careful inspection of the temperature derivative of electrical resistance $dR/dT$ (inset of Fig.1b) - one is at 103 K and the other is at 90 K. Recent experimental results indicate that both structural and AFM phase transitions occur in the parent 10-3-8 compound[3,7,14] and that these transition temperatures can be inferred from the dip temperatures in $dR/dT$,[15] however, no clear evidence shows which one happens first. We therefore denote these two characteristic temperatures in $dR/dT$ by $T_1$ and $T_2$, respectively. The magnetic susceptibility data above 100 K presents only weak temperature dependence, nearly $T$-linear, indicating that the material is in a paramagnetic (PM) metallic-like state at higher temperature (Fig.1c). [1,16-17]

Figure 2 depicts the *in-situ* temperature dependence of resistance measured at different pressures for the 10-3-8 compound. It is apparent that the resistance is reduced over the entire temperature range when pressure is applied (Fig.2a) and that $T_1$ shifts to lower temperature (Fig.2a and inset). The evolution of $T_2$ with pressure is more obscure, so the PM-AFM transition temperature is estimated by assuming that $T_1$ and $T_2$ mirror each other with pressure, as is seen in the undoped FeAs-122 phase. [18-20] At a pressure of ~3.5 GPa, $T_1$ is invisible and a sharp drop in the resistance is seen at ~7 K. The resistance drop becomes more dramatic with increasing pressure (Fig.2b) and zero resistance is achieved in another series of high-pressure resistance measurements (inset of Fig.2b), indicating the existence of superconductivity. The normal-state resistance of the sample shown in the inset of Fig.2b is given in supplementary information (Fig.S1). At pressure above 7 GPa, superconductivity is completely suppressed (Fig.2c). From ambient pressure to 10.8 GPa, the normal-state resistance minimum ($T'$) moves to lower temperature, and the upturn feature in resistance becomes less pronounced. This is similar to what is observed in Pt-doped 10-3-8 superconductors at ambient pressure. [2-5,7,8] In the pressure range of 10.8-24.3 GPa, the normal state resistance minimum disappears and the temperature ($T$) dependent resistance ($R$) shows metallic behavior down to 4 K. $R(T)$ is almost unchanged between 24.3 - 31.4 GPa (Fig.2d).

To provide further evidence that the 10-3-8 phase in the pressure range of 3.5-7

GPa is superconducting, magnetic fields parallel to the *c* axis is applied on the sample at fixed pressure of 5.8 GPa (Fig.3a). A monotonic suppression of the resistance transition is observed with applied fields, *i.e.* the resistance drop moves to lower temperature. To fully characterize the superconducting state in the compressed 10-3-8 phase, we then performed alternating current (*ac*) susceptibility measurements at high pressure in a diamond anvil cell. As shown in Fig.3b, no diamagnetism is detected down to 4 K at 0.2 GPa. However, at 3.5 GPa a remarkable diamagnetism appears at 7.1 K, which unambiguously indicates that the pressure-induced resistance drop in the 10-3-8 phase is a superconducting transition. *Tc* further increases to 7.5 K at 4.2 GPa but it no longer exists at 8.9 GPa, consistent with our resistance data. We estimated the shielding volume fraction by the way of comparing the plunge percentages of diamagnetism in the 10-3-8 sample with that of a "standard sample" ($MgB_2$). We found that the shielding volume fraction of the 10-3-8 sample is about 37%, indicating that the pressure-induced superconductivity has a bulk nature.[21] *In-situ* high pressure synchrotron X-ray diffraction measurements at room temperature found no pressure-induced structural phase transitions under pressure up to ~14 GPa (Supporting Information Fig. S2 and Fig. S3).

Employing our data, we construct the pressure-temperature phase diagram for the 10-3-8 compound, as displayed in Fig.4. $T_1$ is very sensitive to pressure and decreases with increasing pressure at an initial rate of -21K/GPa. At ~ 3.5 GPa, $T_1$ is fully suppressed and superconductivity appears. This is reminiscent of what has been found for LaFeAsO and A-122 (A=Ca, Sr and Ba), in the sense that superconductivity arises in the proximity of a suppressed AFM phase,[1,14,15,18-19]] revealing that the suppression of AFM long-range order is also crucial for achieving superconductivity in the 10-3-8 phase. The superconducting dome, with a maximum *Tc* of ~8.5 K at 4.1 GPa, lies in the pressure range from 3.5 GPa to 7 GPa. Furthermore, the crossover temperature (*T'*) of the semiconducting-to-metallic like state is reduced under pressure. The superconducting dome stands within the semiconducting normal state region, a characteristic that has also been reported in the 10-3-8 superconductor at ambient pressure as a function of Pt doping.[3-5,7,8] The AFM transition vanishes at the almost same temperature in the phase diagram for both pressurized and doped systems, and simultaneously their superconductivity emerges at low temperature with nearly the same *Tc*, which motivates a comparison between the two phase diagrams, as has for example been done for pressurized and doped $V_2O_3$,[22] with the expectation of unveiling hidden relations between pressure-induced and doping-induced superconductivity in the iron pnictides.

Taking the pressure and doping values where the AFM transition vanishes as the scaling points for merging the pressure-temperature and doping-temperature phase diagrams, we add the temperature-composition electronic phase diagram of the Pt-doped 10-3-8 phase to the same figure (Fig.4). It can be seen that the doping composition is scaled by 0.018 Pt/2.24 GPa and also that the points presenting the two $T_1$ phase match each other, *i.e,* the rate of suppressing $T_1$ is nearly same for pressurized and doped systems. Interestingly, there is no coexistence region of AFM phase and superconducting phase in the phase diagram for either pressurized or Pt-doped 10-3-8 samples, which is completely different from that what is commonly seen in pressurized and doped iron pnictide superconductors with 122 type.[12,13]

At pressures below ~4 GPa, the effect of pressure on developing and stabilizing superconductivity is found to be similar to that of doping effects below x~0.03, from which we find that the tuning effect of 1 GPa is equivalent to 0.008 Pt doping. This can be explained by the fact that applied pressure increases the bandwidths in the 10-3-8 phase, as has been seen in other materials. [23,24] The decreasing semiconducting behavior of the 10-3-8 phase with increasing applied pressure may reflect that the gap in the $Pt_3As_8$ layer is closing due to the increase in bandwidth. As a result of the increased bandwidth in the $Pt_3As_8$ layer under pressure, some formerly full states are pushed above Fermi energy and charge transfer occurs between the $Pt_3As_8$ layer and the $Fe_2As_2$ layer, leading to the electronic doping of the $Fe_2As_2$ layers. This hypothesis is supported by our Hall measurements for the undoped 10-3-8 phase under pressure below 4.8 GPa, above the pressure of which the Hall resistance of the sample is not detectable due to the pressure-induced microscopic cracks. The Hall coefficient ($R_H$) obtained at different pressures, shown in Fig. 5, is negative at all pressure points investigated. Within a single band, the carrier density $n$ can be estimated as $-1/eR_H$. $n$ (at 20 K) against pressure is plotted in inset of Fig.5, together with $n$ against Pt doping. It is seen that $n$ increases from $3.1\times 10^{19}$ /cm$^3$ at 0.2 GPa to $11.3\times 10^{19}$ /cm$^3$ at 3.5 GPa, indicating that the pressure makes the sample more $n$ type. The pressure-induced change in carrier concentration in the pressure region <4 GPa is similar and comparable to that seen for Pt-doping,[8] undisputedly confirming that pressure plays the same role as electron doping in the range of moderate pressure. On the basis of the scaled phase diagram, we can propose that the critical carrier concentration ($n_c$) for pressure-induced superconductivity in the undoped 10-3-8 phase is about $11.3\times 10^{19}/cm^3$. This correlation, achieved by scaling analysis, is valuable to further studies on the mechanism of superconductivity in iron-based superconductors from theoretical side.

Unexpectedly, increasing pressure above 4 GPa the *Tc* of the 10-3-8 phase decreases (Fig.4), while the *Tc* values at the corresponding scaled Pt doping levels continue to increase, which implies that the pressure-induced change in lattice structure is non-negligible. The breakdown of the pressure-temperature scaling in higher pressure and heavy doping regime implies that pressure takes a different path in tuning superconductivity from that of Pt-doping in the 10-3-8 phase. This may reveal that there exist alternative microphysical mechanisms in impacting on the Fermi surface topology (FST) for both cases. Since the FST of the iron pnictides is thought to play a key role in determining *Tc* [25] but there is no available experimental way to detect the change in FST at high pressure, pressure/doping scaling analysis provides a pathway to study the FST under pressure. This new system for studying the FST under pressure may help to shed insight on the underling mechanism of superconductivity in iron-based arsenide superconductors.

In conclusion, we find the pressure-induced superconductivity in $Ca_{10}(Pt_3As_8)(Fe_2As_2)_5$, a compound of significant current interest, which is an antiferromagnetic semiconductor at ambient pressure but can be a superconductor upon doping with Pt on the Fe site, by *in-situ* high-pressure resistance and magnetic susceptibility measurements. High-pressure Hall coefficient measurements indicate that the pressure induces charge transfer from the charge reservoir to the iron arsenide layers and thus leads to superconductivity. However, scaling of the pressure-induced and doping-induced superconductivity shows that the electronic phase diagrams of the pressurized and chemically-doped 10-3-8 compound are similar in the moderate pressure and doping range but are disparate at higher pressure and heavy doping. These results may bring new insight on the underlying mechanism of superconductivity in the iron-based superconducting materials.

**Experimental section**

Single crystals of Pt-doped and undoped 10-3-8 were grown using the flux method as described in Ref.2. Ambient-pressure electrical resistance and magnetic susceptibility as a function of temperature were performed in a Quantum Design Physical Property Measurement System (PPMS) and a Quantum Design Magnetic Property Measurement System, respectively. Pressure was generated by a diamond anvil cell with two diamond anvils sitting on a Be-Cu supporting plate oppositely. The anvil diameter is about 300 μm. The nonmagnetic rhenium gasket was preindented down to 50μm thickness for different runs. The 10-3-8 single crystal with dimensions of around 80×80×20 μm was loaded into the gasket hole. High-pressure electrical resistance experiments were carried out using a standard four-probe technique. High-pressure alternating current (*ac*) susceptibility measurements were conducted using home-made coils around a diamond anvil [26-28,]. High-pressure Hall coefficient measurements with a four-wire symmetry were carried out for the undoped 10-3-8 phase, with a magnetic field perpendicular to the *ab* plane of the sample to 0.2 Tesla by sweeping the magnetic field at fixed temperature for each given pressure. High-pressure angle dispersive X-ray diffraction (XRD) measurements were carried out at beamline 15U at the Shanghai Synchrotron Radiation Facility (SSRF). The pressure transmitting medium (PTM) used for high-pressure transport measurements is NaCl powder, while the PTM used for high-pressure X-ray diffraction measurements is silicone oil with viscosity of 1 cst. The level of hydrostatic pressure applied on the 10-3-8 phase was confirmed to be hydrostatic by comparing with the data of the $SrFe_2As_2$.[29] Pressure was determined by ruby fluorescence.[30]

**Supporting Information**

Supporting Information is available from the Wiley Online Libarary.

**Acknowledgements**

This work in China was supported by the NSCF (Grant No. 11074294 and 11204059), 973 projects (Grant No. 2011CBA00100 and 2010CB923000) and Chinese Academy of Sciences. The work in the USA has been supported by the AFOSR MURI on superconductivity, grant FA9550-09-1-0593. Dr. Ni acknowledges the Marie Curie Fellowship at Los Alamos National Laboratory.

## References

1. Y. Kamihara, T. Watanabe, M. Hirano, and H. Hosono, *J. Am. Chem. Soc.* **2008**, *130*, 3296-3297.
2. N. Ni, J. M. Allred, B. C. Chan, R. J. Cava, *Proc. Natl. Acad. Sci.* **2011**, *108*, 1019-1026.
3. K. Cho1, M. A. Tanatar1, H. Kim, W. E. Straszheim1, N. Ni, R. J. Cava, R. Prozorov, *Phys. Rev. B* **2012**, *85*, 020504(R).
4. M. Nohara, S. Kakiya, K. Kudo, Y. Oshiro; S. Araki, T. C. Kobayashi, K. Oku, E. Nishibori, H. Sawa, *Solid State Commun.* **2012**, *152*, 635-639.
5. S. Kakiya, K. Kudo, Y. Nishikubo, K. Oku, E. Nishibori, H. Sawa, T. Yamamoto, T. Nozaka, M. Nohara, *J. Phys. Soc. Jpn.* **2011**, *80*, 093704 .
6. T. Sturzer, G. Derondeau, D. Johrendt, *Phys. Rev. B* **2012**, *86*, 060516(R).
7. C. Löhnert, T. Stürzer, M. Tegel, R. Frankovsky, G. Friederichs, D. Johrendt, *Angew. Chem. Int. Ed.* **2011**, *50*, 9195-9199.
8. Z. J. Xiang, X. G. Luo, J. J. Ying, X. F. Wang, Y. J. Yan, A. F. Wang, P. Cheng, G. J. Ye, X. H. Chen, *Phys. Rev. B* **2012**, *85*, 224527.
9. J. Zhao, Q. Huang, C. de la Cruz, S. L. Li, J. W. Lynn, Y. Chen, M. A. Green, G. F. Chen, G. Li, Z. Li, J. L. Luo, N. L. Wang, P. C. Dai, *Nat. Mat.* **2008**, *7*, 953-959.
10. J. H. Dai, Q. M. Si, J. X. Zhu, E. Abrahams, *Proc. Natl. Acad. Sci. USA* **2009**, *106*, 4118-4121.
11. K. Jin, N. P. Butch, K. Kirshenbaum, R. L. Greene, *Nature* **2011**, *476*, 73-75.

12. X. Y. Zhu, F. Han, G. Mu, J. Tang, J. Ju, K. Tanigaki, H. H. Wen, *Phys. Rev. B* **2010**, *81*, 104525.

13. J. Paglione, R. L. Greene, *Nat. Phys.* **2010**, *6*, 645-658.

14. T. Zhou, G. Koutroulakis, J. Lodico, N. Ni, J. D. Thompson, R. J. Cava, S. E. Brown, *J. Phys.: Condens. Matter* **2013**, *25*, 122201.

15. N. Ni, W. E. Straszheim, D. J. Williams, M. A. Tanatar, R. Prozorov, E. D. Bauer, F. Ronning, J. D. Thompson, R. J. Cava, *Phys. Rev. B* **2013**, *87*, 060507.

16. M. S. Torikachvili, S. L. Bud'ko, N. Ni, P. C. Canfield, *Phys. Rev. Lett.* **2008**, *101*, 057006.

17. S. H. Baek, H. Lee, S. E. Brown, N. J. Curro, E. D. Bauer, F. Ronning, T. Park, J. D. Thompson, *Phys. Rev. Lett.* **2009**, *102*, 227601.

18. H. Lee, E. Park, T. Park, V. A. Sidorov, F. Ronning, E. D. Bauer, J. D. Thompson, *Phys. Rev. B* **2009**, *80*, 024519.

19. K. Kitagawa, N. Katayama, H. Gotou, T. Yagi, K. Ohgushi, T. Matsumoto, Y. Uwatoko, and M. Takigawa, *Phys. Rev. Lett.* **2009**, *103*, 257002.

20. S. A. J. Kimber, A. Kreyssig, Y. Z. Zhang, H. O. Jeschke, R. Valentí, F. Yokaichiya, E. Colombier, J. Q. Yan, T. C. Hansen5, T. Chatterji, R. J. McQueeney, P. C. Canfield, A. I. Goldman, D. N. Argyriou, *Nat. Mater.* **2009**, *8*, 471-475.

21. H. Lee, E. Park, T. Park, V.A. Sidorov, F. Ronning, E. D. Bauer, J.D. Thompson, *Phys. Rev. B* **2009**, *80*, 024519.

22. D. B. McWhan, A. Menth, J. P. Remeika, W. F. Brinkman, T. M. Rice, *Phys. Rev.*

***B*** **1973**, *7*, 1920.

23. M. Imada, A. Fujimori, Y. Tokura, *Rev. Mod. Phys*. **1998**, *70*, 1039.
24. P. Limelette, A. Georges, D. Jérome, P. Wzietek, P. Metcalf, J. M. Honig, *Science* **2003**, *302*, 89.
25. C. Liu, A. D. Palczewski, R. S. Dhaka, Takeshi Kondo, R. M. Fernandes, E. D. Mun, H. Hodovanets, A. N. Thaler, J. Schmalian, S. L. Bud'ko, P. C. Canfield, A. Kaminski, *Phys. Rev. B* **2011**, *84*, 020509(R).
26. L. L. Sun, X. J. Chen, J. Guo, P. W. Gao, Q. Z. Huang4, H. D. Wang, M. H. Fang5, X. L. Chen, G. F. Chen, Q. Wu, C. Zhang1, D. C. Gu, X. L. Dong1, L. Wang, K. Yang7, A. G. Li, X. Dai, H. K. Mao, Z. X. Zhao, *Nature* **2012**, *483*, 67-69.
27. M. Debessai, T. Matsuoka, J. J. Hamlin, J. S. Schilling, *Phys. Rev. Lett.* **2009**, *102*, 197002.
28. T. Tomita, S. Deemyad, J. J. Hamlin, J. S. Schilling, V. G. Tissen, B. W. Veal, L. Chen, H. Claus, *J. Phys.: Condens. Matter* **2005**, *17*, S921-S928.
29. K. Matsubayashi, N. Katayama, K. Ohgushi, A. Yamada, K. Munakata, T. Matsumoto, Y. Uwatoko, *J. Phys. Soc. Jpn.* **2009**, *78*, 073706.
30. H. K. Mao, J. Xu, P. M. Bell, *J. Geophys. Res.* **1986**, *91*, 4673-4676.
31. N. Ni and R. J. Cava, unpublished (Ni, do you published these data? If yes, would you pls add the name of the journal?)

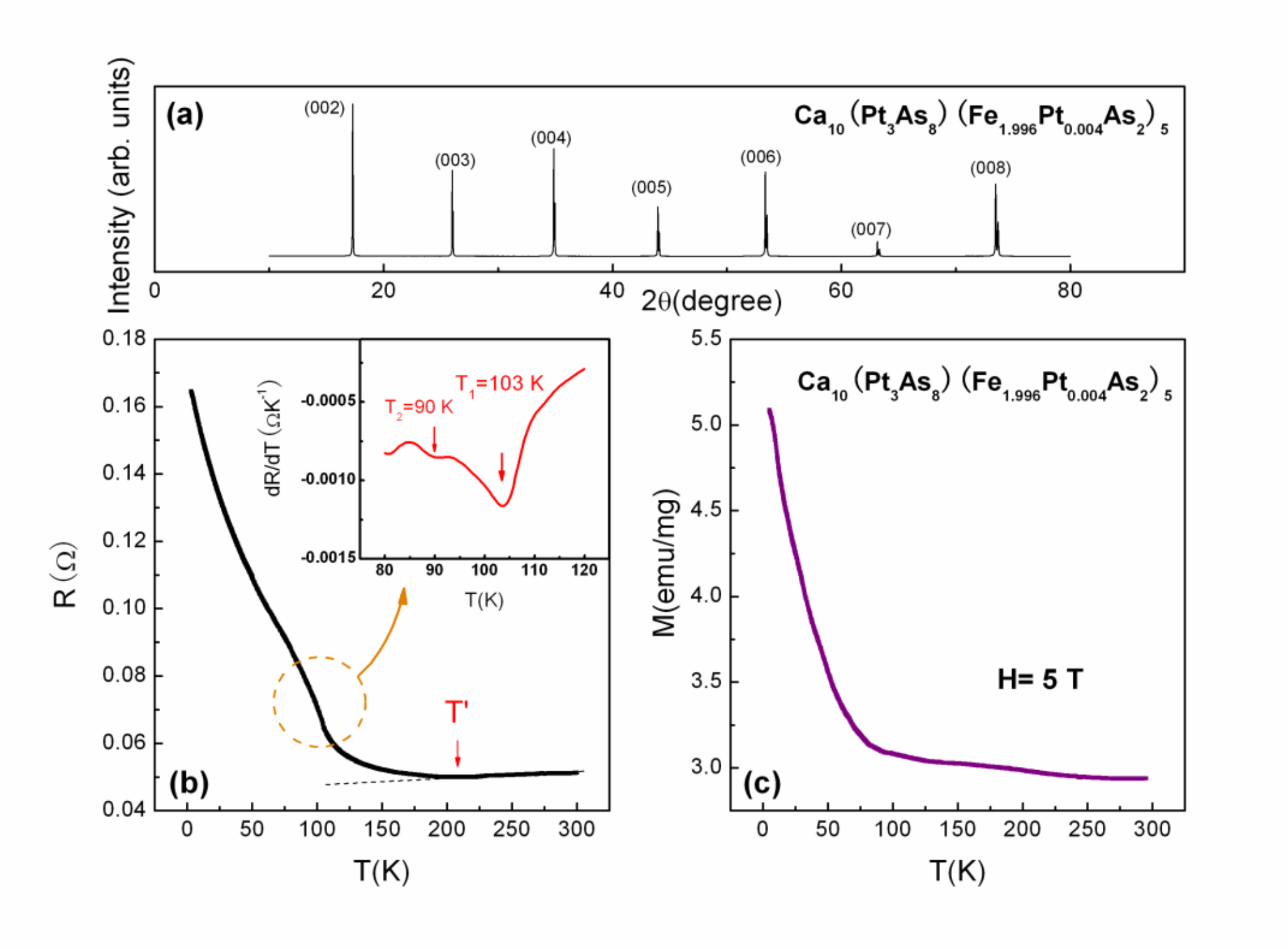

**Figure 1 X-ray diffraction pattern of the 10-3-8 single crystal and temperature dependence of resistance and susceptibility at ambient pressure.** (a) X-ray diffraction pattern collected at room temperature and ambient pressure for the sample with actual composition of $Ca_{10}(Pt_3As_8)(Fe_{1.996}Pt_{0.004}As_2)_5$. (b) Electrical resistance as a function of temperature for the same single crystal. The *T′* represents the crossover temperature from metallic state to semiconducting-like state. The Inset displays the temperature derivative of electrical resistance *dR/dT*. The dip at 103 K ($T_1$) is related to a structural phase transition and the dip at 90 K ($T_2$) is associated with the AFM transition. (c) Temperature dependence of magnetic susceptibility for the sample cut from the same batch for the resistance measurement.

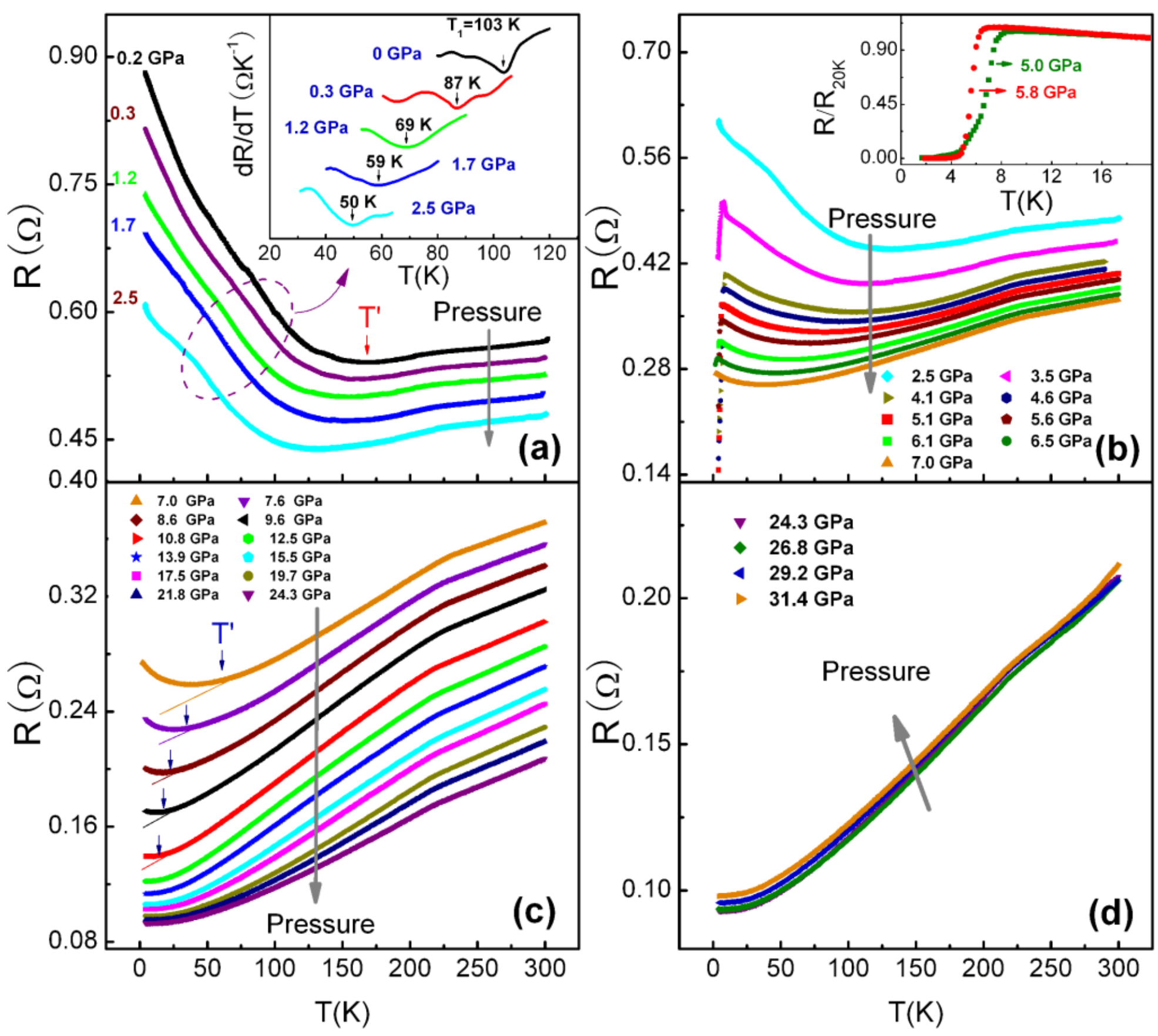

**Figure 2 Temperature dependence of electrical resistance for the 10-3-8 phase at different pressures.** (a) Resistance-temperature (R-T) curves in the semiconducting-like state under pressure to 2.5 GPa. The inset demonstrates the temperature derivative of electrical resistance *dR/dT* obtained at different pressures. $T_1$ and $T_2$ shift to lower temperatures. (b) A resistance drop is observed in the pressure range between 3.5 GPa - 7 GPa. Inset shows the R-T curves measured at different pressures in another run, displaying zero resistance at low temperature. (c) The semiconducting-like state at lower temperatures is suppressed upon increasing pressure. (d) The resistance-temperature curve exhibits metallic behavior at pressure

above 24.3 GPa.

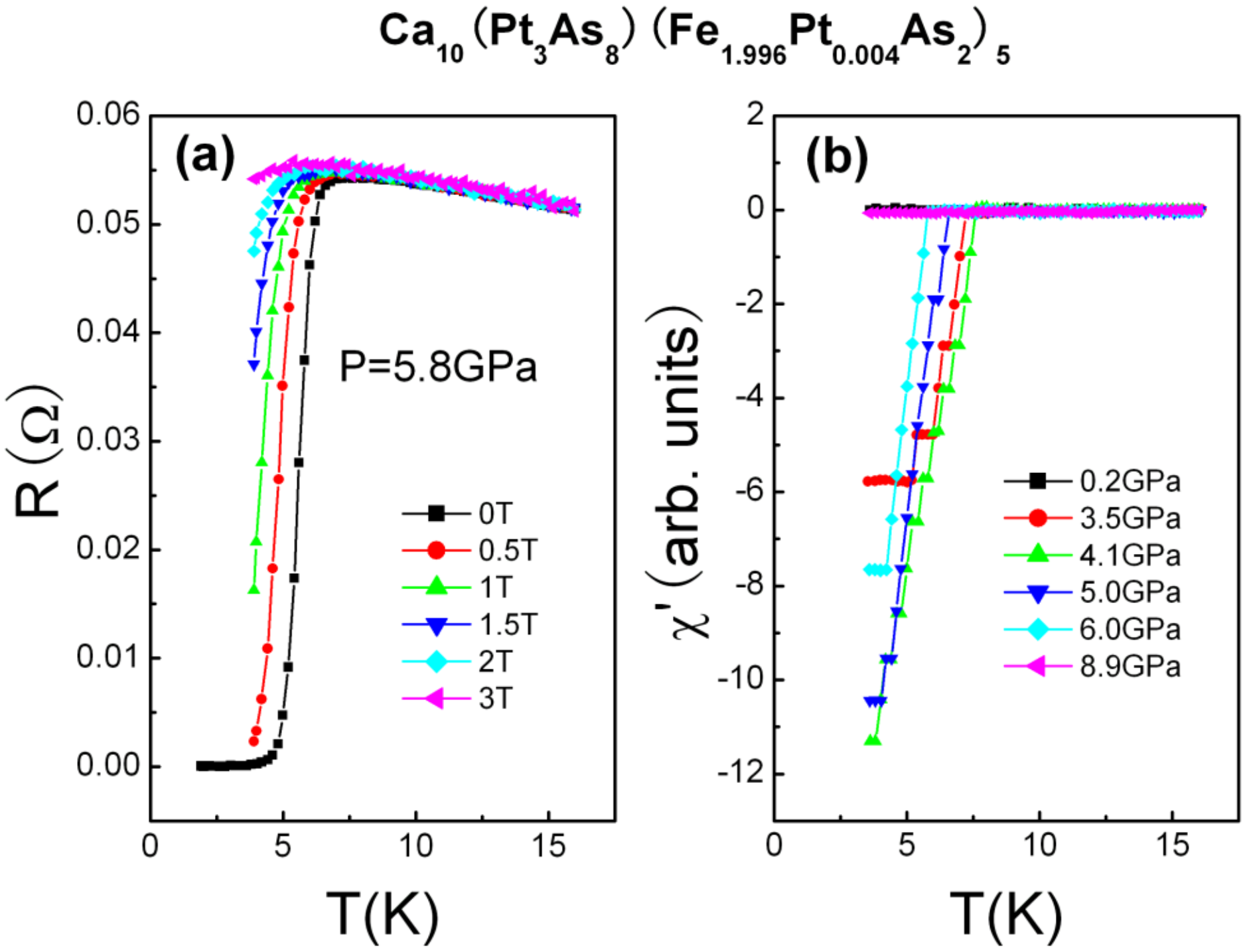

**Figure 3 Electrical resistance as a function of temperature measured at different magnetic fields at a fixed pressure of 5.8 GPa, and high-pressure alternating current (*ac*) susceptibility measurements at different pressures for the 10-3-8 single crystal.** (a) The resistance-temperature curve of the sample at different magnetic fields. The resistance drop is shifted to lower temperature and suppressed with increasing field. (b) Temperature dependence of the real part of the *ac* susceptibility at different pressures. Diamagnetism starting at 3.5 GPa is clearly seen, demonstrating the presence of a superconducting state.

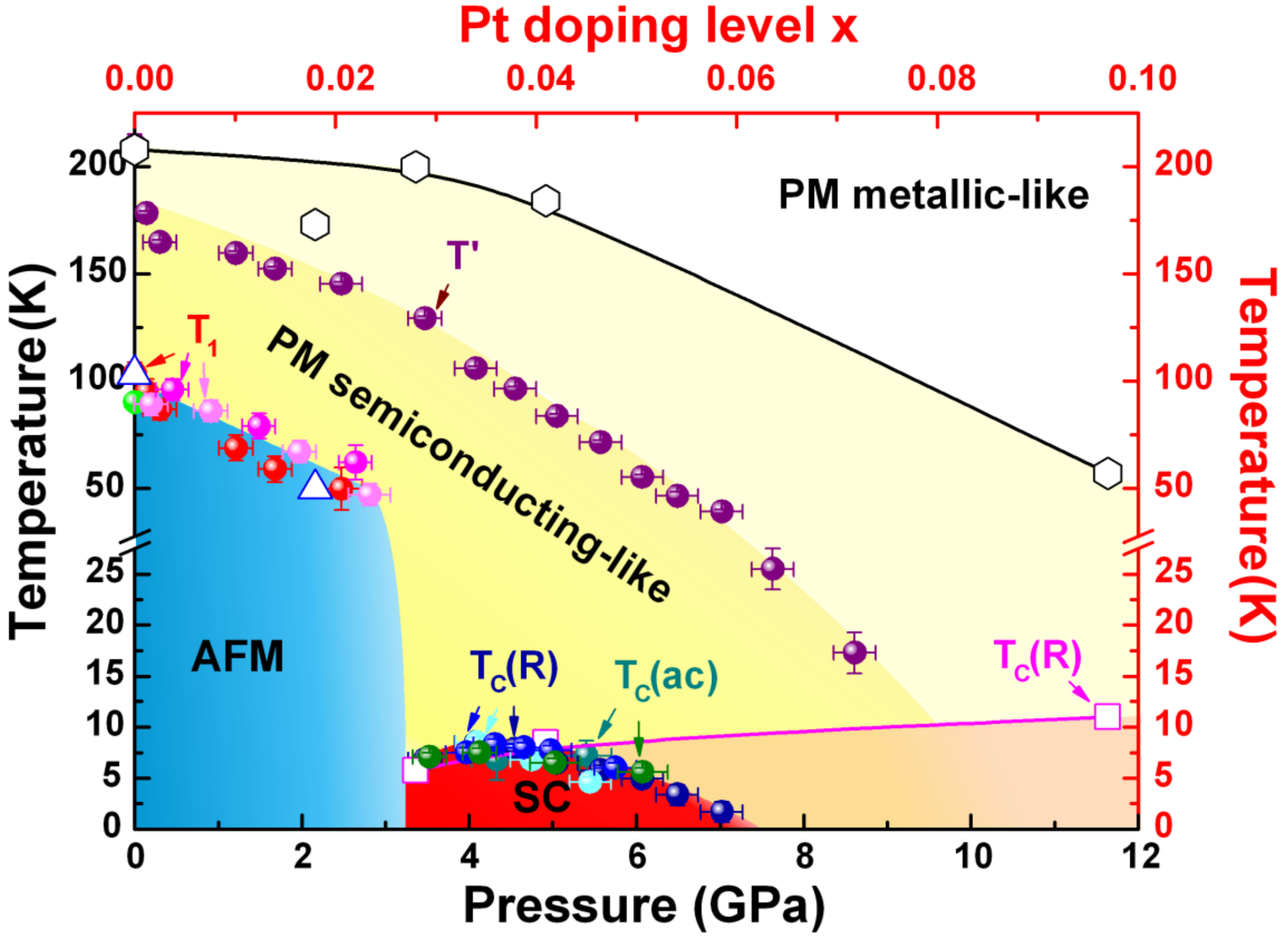

**Figure 4 Temperature-pressure electronic phase diagram for the 10-3-8 phase, and scaled temperature-doping phase diagram, showing their equivalence in the range of moderate pressure and light doping and nonequivalence in the range of higher pressure and heavy doping.** Open symbols is the corresponding data from the temperature-doping phase diagram of Pt-doped 10-3-8.[31] The navy and cyan solid circles represent the *Tc* values obtained from high-pressure resistance (*Tc(R)*) and the dark cyan solid circle repensents the *Tc* value measured from high-pressure *ac* susceptibility (*Tc(ac)*) measurements. The purple solid circle reprents the crossover temperature (*T'*) from metallic state to semiconducting-like state. The red and pink solid circles show the AFM transition temperature determined by the temperature derivative of electrical resistance *dR/dT* obtained at different pressures. SC represents the superconducting region. AFM represents the

antiferromagnetic phase. PM represents the paramagnetic phase.

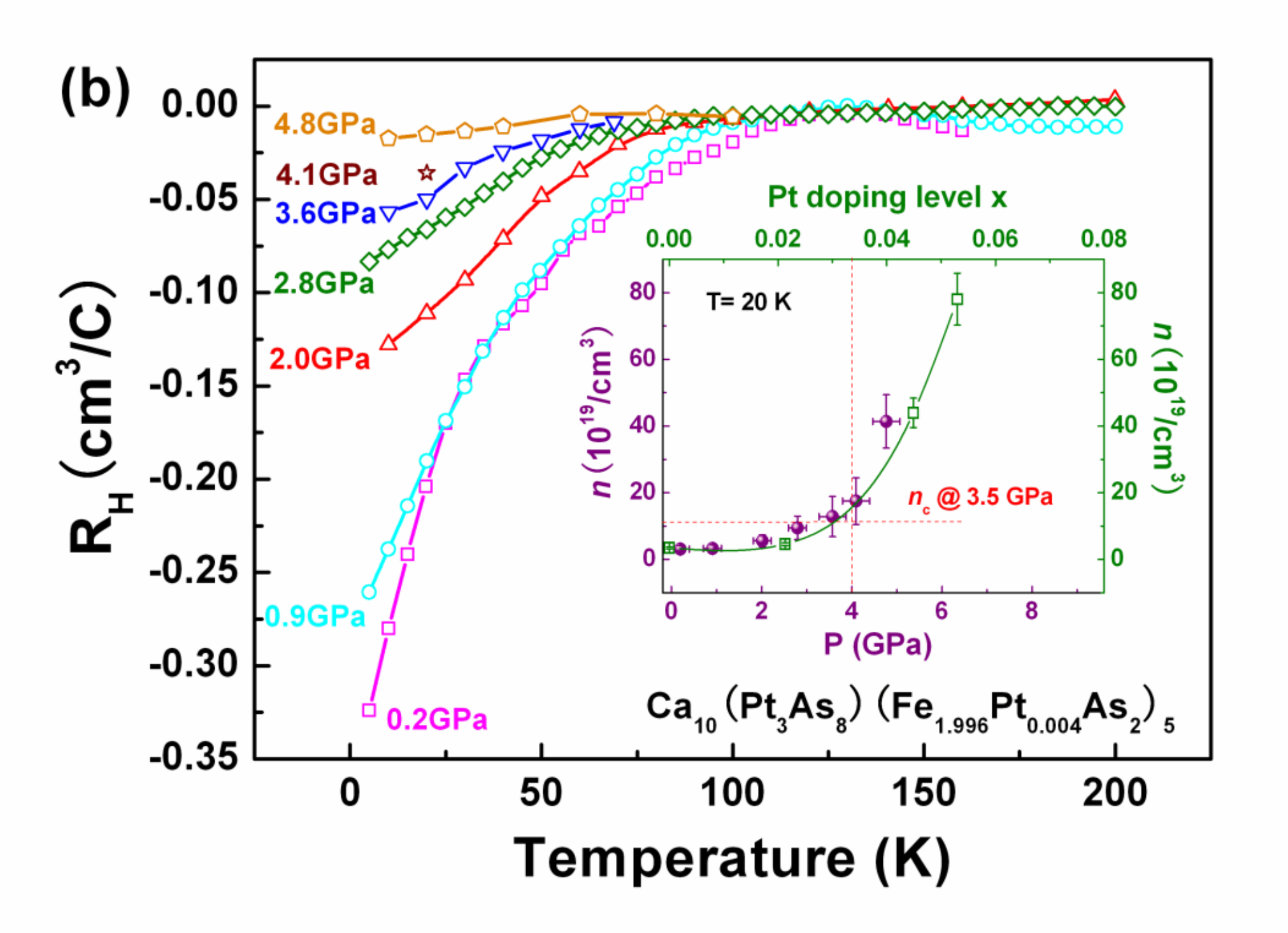

**Figure 5 Hall coefficient as a function of temperature obtained at different pressures and pressure dependent of carrier concentration (*n*) for the undoped 10-3-8 phase (inset).** The negative Hall coefficients in the undoped 10-3-8 at different pressures demonstrate that the carriers are electron dominated. The carrier concentration as a function of pressure and its evolution with pressure below ~4 GPa is nearly the same as that of Pt doping below ~0.03, while disparate from pressure above ~4 GPa. The data of *n* for Pt doping shown in inset are taken from Ref. [8]. The critical carrier concentration for occurrence of superconductivity in both pressurized and Pt-doped 10-3-8 is determined to be ~11× $10^{19}$ /$cm^3$.